\documentclass[preprint]{./acm_proc_article-sp}
\usepackage{epsfig}
\usepackage[square,comma,numbers,sort&compress]{natbib}
\usepackage{multirow}

\newcommand{\UPLB}{University of the Philippines Los Ba\~{n}os}

\newcommand{\Papers}{\mathcal P}
\newcommand{\Authors}{\mathcal A}
\newcommand{\E}{\mathcal E}
\newcommand{\AuthorsPerPaper}{A_\Papers}
\newcommand{\PapersPerAuthor}{P_\Authors}
\newcommand{\CollabsPerAuthor}{C_\Authors}
\newcommand{\MinAuthorsPerPaper}{A_{\Papers\mathrm{MIN}}}
\newcommand{\AvgAuthorsPerPaper}{A_{\Papers\mathrm{AVG}}}
\newcommand{\MaxAuthorsPerPaper}{A_{\Papers\mathrm{MAX}}}
\newcommand{\MinPapersPerAuthor}{P_{\Authors\mathrm{MIN}}}
\newcommand{\AvgPapersPerAuthor}{P_{\Authors\mathrm{AVG}}}
\newcommand{\MaxPapersPerAuthor}{P_{\Authors\mathrm{MAX}}}
\newcommand{\MinCollabsPerAuthor}{C_{\Authors\mathrm{MIN}}}
\newcommand{\AvgCollabsPerAuthor}{C_{\Authors\mathrm{AVG}}}
\newcommand{\MaxCollabsPerAuthor}{C_{\Authors\mathrm{MAX}}}
\newcommand{\PAG}{\mathbf{PAG}}
\newcommand{\PAGS}{\mathrm{PAG}}
\newcommand{\CAG}{\mathbf{CAG}}
\newcommand{\CAGS}{\mathrm{CAG}}
\newcommand{\CAAG}{\mathbf{CAAG}}
\newcommand{\PAM}{\mathbf{PAM}}
\newcommand{\CAM}{\mathbf{CAM}}
\newcommand{\CAAM}{\mathbf{CAAM}}
\newcommand{\power}{\varphi}
\newcommand{\union}{\bigcup}
\newcommand{\intersection}{\bigcap}
\newcommand{\degree}{\Delta}
\newcommand{\degreeP}{\degree^\Papers}
\newcommand{\degreeA}{\degree^\Authors}
\newcommand{\degreeC}{\degree^C}

\begin{document}
\title{Authorship Patterns in\\ Computer Science Research in the Philippines}
\numberofauthors{1}
\author{
\alignauthor Jaderick P. Pabico\titlenote{\tt http://www.ics.uplb.edu.ph/jppabico}\\
   \affaddr{Institute of Computer Science}\\
   \affaddr{\UPLB}\\
   \affaddr{College 4031, Laguna}\\
   \email{jppabico@uplb.edu.ph}
}
\date{}
\maketitle
\begin{abstract}
We studied patterns of authorship in computer science~(CS) research in the Philippines by using data mining and graph theory techniques on archives of scientific papers presented in the Philippine Computer Science Congresses from 2000 to 2010 involving 326~papers written by 605~authors. We inferred from these archives various graphs namely, a paper--author bipartite graph, a co-authorship graph, and two mixing graphs. Our results show that the scientific articles by Filipino computer scientists were generated at a rate of 33~papers per year, while the papers were written by an average of 2.64~authors (maximum=13). The frequency distribution of the number of authors per paper follows a power-law with a power of $\power=-2.04$ ($R^2=0.71$). The number of Filipino CS researchers increases at an annual rate of 60~new scientists. The researchers have written an average of 1.42~papers (maximum=20) and have collaborated with 3.70~other computer scientists (maximum=54). The frequency distribution of the number of papers per author follows a power law with $\power=-1.88$ ($R^2=0.83$). This distribution closely agrees with Lotka's {\em law of scientific productivity} having $\power\approx -2$. The number of co-authors per author also follows a power-law with $\power=-1.65$ ($R^2=0.80$). These results suggest that most CS~papers in the country were written by scientists who prefer to work alone or at most in small groups. These also suggest that few papers were written by scientists who were involved in large collaboration efforts. The productivity of the Philippines' CS researchers, as measured by their number of papers, is positively correlated with their participation in collaborative research efforts, as measured by their number of co-authors (Pearson $r=0.7425$). The Filipino CS scientists follow a low dissortative mixing when choosing a collaborator either in terms of the collaborator's number of papers ($r=-0.1015$), or its number of co-authors ($r=-0.0398$). This means that a Filipino CS researcher with high numbers of papers and co-authors chooses a collaborator whose numbers of papers and co-authors are low.
\end{abstract}

\section{Introduction}
The patterns of authorship of scientific research articles reflect how the volume of knowledge was generated by the scientists in a country. The number of quality papers that a nation's researchers write within a time period reflects the scientific productivity of that nation's scientists. The number of authors who wrote a particular research article, on the other hand, mirrors the number of manpower needed to generate the knowledge embodied in the paper. The number of co-authors that a scientist has tells the participation of that scientist in collaborative research efforts, as well as that scientist's dependency with other researchers to generate knowledge. This paper presents the authorship patterns of computer science~(CS) research in the Philippines as induced from the archives of scientific papers presented in the Philippine Computer Science Congresses (PCSC) from 2000 to 2010.\footnote{The PCSC started in 2000 but the 2001 papers are not accessible to the author. There was no PCSC conducted in 2002~\citep{adorna10}.} Although the subject of this paper falls under the CS subdisciplines of graph theory, data structures, information retrieval and mining, visualization, and pattern discovery, the subject matter will be of more interest to the whole computing science community in the Philippines for just one reason: {\em it is all about the Filipino computer scientists}. We hope that with this paper, we can understand several factors in CS~research that are unique in the Philippine setting. For example, we can quantify the bounds of the amount of scientific knowledge that the Filipino computer scientists generated, as well as the bounds of the number of Filipino computing scientists who conducted research in the past years. We can also identify who are the most prolific computer scientists, as well as those who are with the most number of research collaborators. In general, understanding the patterns on how the Filipino computer scientists generate knowledge may provide discernment on information breakdowns, bottlenecks, and structural holes in the scientific community of CS in the Philippines.

In recent years, the advent of advanced computer-based archiving technologies made most scientific works in the last 10~to 50~years easily accessible via any digital media by virtually anyone from anywhere. Examples of such archives are the Los Alamos e-Print Archive (LAePA)~\citep{losalamos}, the Medline Database (Medline)~\citep{medline}, the Standford Public Information Retrieval System (SPIRES)~\citep{spires}, the Network of Computer Science Technical Reference (NCSTRL)~\citep{ncstrl}, the DBLP Computer Science Bibliography (DBLP)~\citep{dblp}, the Samahang Pisika ng Pilipinas (~SPP)~\citep{spp}, the Transactions of the National Academy of Science and Technology-Annual Scientific Meetings (NAST-ASM), and the Proceedings of the Philippine Society of Agricultural and Biosystems Engineering (BAE)~\citep{conf07,conf08,conf09} (Table~\ref{tab:archives}). These archives compile scientific papers that were recently used by some researchers~\citep{newman01a,newman01b,barabasi02,bird09,pabico09a,pabico09b,villanueva09,pabico10} who conducted data mining techniques to understand the complex nature in scientific research in various fields. Inferred from these archives are results that show that the average papers per author ranges from~2 to~7, while the papers were written by an average of~2 to~9 authors. Depending on the scientific discipline, a given author has an average of~3 to~173 collaborators~\citep{newman01a,barabasi02,pabico09a,pabico09b,pabico10}. 

\begin{table}[htb]
   \caption{List of example scientific paper archives as used by various researchers~\citep{newman01a,newman01b,barabasi02,bird09,pabico09a, pabico09b,villanueva09,pabico10}. The number of papers and of authors are the numbers when the studies were conducted by the respective authors.}\label{tab:archives}
   \centering\begin{tabular}{lcrr}
    \hline\hline
	Archive &  Year   & Number of & Number of\\
	Name    & Started & Papers    & Authors\\
	\hline
	\multicolumn{4}{l}{\underline{International Archives}}\\
	\quad LAePA       & 1992 & $>161,000$ & $>94,000$\\
	\quad Medline     & 1961 & $>216,000$ & $>152,000$\\
	\quad SPIRES      & 1990 & $\approx 66,000$ & $>56,000$\\
	\quad NCSTRL      & 1974 & $\approx 13,000$ & $\approx 12,000$\\
	\quad DBLP        & 1960 & $\approx 84,000$ & $\approx 95,000$\\
	\multicolumn{4}{l}{\underline{Philippine-based Archives}}\\
	\quad SPP         & 2001 & $699$ & $840$ \\
    \quad Agriculture & 2006 & $235$ & $645$\\
	\quad NAST-ASM    & 2006 & $720$ & $1,780$\\
	\quad ABE         & 2007 & $90$  & $171$\\ 
	\hline\hline
   \end{tabular}   
\end{table}

In the Philippines, we have previously utilized the archives of scientific posters presented at the recent NAST-ASM in an initial attempt to understand the authorship patterns of Filipino agricultural scientists~\citep{pabico09a}. Although Philippine-based scientific journals and proceedings in agricultural science abound, we assumed that the papers compiled in the NAST-ASM archives represent the majority of scientific knowledge discovered by Filipino agricultural scientists, not only because of the sheer volume of knowledge it contains, but also because of the quality of knowledge presented having been reviewed, and often times authored, by no less than the nation's Academicians and National Scientists. The papers used in this study were categorized under the Agricultural Sciences Division (ASD) and involved 235 poster abstracts written by 645 authors spanning the recent four years from 2006 to 2009. In this study, we have found that the Filipino agricultural scientists have written an average of 1.39 papers (maximum=13), while they have collaborated with an average of 2.70 scientists (maximum=28). Their research papers were written by an average of 3.81 Filipino authors (maximum=15).

Using the same NAST-ASM archives, we recently expanded~\citep{pabico09b} the above study to involve all six NAST scientific divisions encompassing various scientific disciplines: The ASD; The biological sciences (BSD); The chemical, mathematical and physical sciences (CMPSD); The engineering sciences and technology (ESTD); The health sciences (HSD); And the social sciences (SSD). This expanded study involved 720 papers written by 1,780 authors. Because of the sheer volume of scientific discoveries contained in the archive, we assumed that the papers represent the major scientific work of Filipino scientists in various disciplines in the past four years from 2006 to 2009. Again, our previous assumption holds that the archives not only contain high quantity of scientific discoveries in the Philippines, but more importantly, high quality research results for the same reason as mentioned above. The results of our analysis show that the Filipino scientists have written an average of 1.52 papers (maximum=40), while they have collaborated with an average of 2.82 scientists (maximum=66). The scientific papers have been written by an average of 3.70 Filipino authors (maximum=22).

Using the NAST-ASM archives to infer the authorship patterns of scientists from specific disciplines proved to be difficult to do, even though works of scientists in a specific field might already be included in the archives. Examples of such disciplines are the Physics, the Agricultural and Biosystems Engineering (ABE), and the CS disciplines.  The reason for this is that the NAST-ASM archives did not label both the scientists and the research works as belonging to either the Physics, the ABE or the CS discipline. In fact, Physics papers are classified only under CMPSD, while ABE papers may be classified within two out of six NAST divisions namely, ASD and ESTD. Both ASD and ESTD involved papers from various other fields that are not ABE in nature, such as entomolgy, biochemistry, forestry, information technology and all other engineering fields. Researches from the CS discipline, on the other hand, maybe classified under CMPSD and ESTD, which also involved various other fields that are not CS in nature. Thus, to analyze the authorship patterns of Filipino scientists and researchers in specific disciplines, separate archives must be used to better reflect the works and workers in the said discipline. In the case of Physics and ABE, their respective archives actually exist as the Proceedings of the Samahang Pisika ng Pilipinas~\citep{spp}, and the Proceedings of the Joint International Agricultural Engineering Conference and Exhibition of the Philippine Society of Agricultural Engineers (PSAE)~\citep{conf07,conf08,conf09}. Both proceedings are archived in digital format. We have analyzed the authorship patterns of Filipino physicists from 2001 to 2005 involving 699~papers written by~840 authors~\citep{villanueva09}, as well as those of the ABE scientists over the recent 3-year period from 2007 to 2009 involving 90~papers written by~171 authors~\citep{pabico10}. Our results in these studies are summarized in Table~\ref{tab:all-collabs-summary}, together with the summary of the previous works discussed above for comparison purposes.

In this current effort, to understand the authorship patterns of Filipino CS researchers, we have applied data mining and graph theory techniques on archives of papers presented in the PCSC~\citep{pcsc00, pcsc03,pcsc04,pcsc05,pcsc06,pcsc07,pcsc08,pcsc09,pcsc10} from 2000 to 2010. The 9--year archive has accumulated 326~papers written by 605~authors. We have found out, on the average, that the CS research papers were authored by 2.64 Filipino scientists, while the CS researchers themselves have written 1.42 papers and have collaborated with 3.70 other scientists. Aside from computing these fundamental quantities to compare the CS community with other scientific disciplines in the country, we also computed the respective frequency distributions of these quantities. The power law nature of these distributions suggest that most CS~papers were authored by those who have a few collaborators, while few of the papers were authored by those who have a large list of collaborators. We have found a statistical evidence suggesting that the productivity of computer scientists in the country is positively correlated with the scientists' participation in a number of collaborative research endeavors. We have also observed low assortative mixing among authors when choosing a collaborators in terms of the collaborator's scientific productivity, as well as the collaborator's number of collaborators. We hope that the results contributed by this paper could later be used to aid the various stakeholders (e.g., funding agencies and professional organizations) in providing opportunities to accelerate knowledge generation in the field of CS in the country, as well as in strengthening the efficiency and effectiveness of existing formal research and technical communication channels.
\begin{table*}[htb]
\caption{Fundamental statistics of various different scientific collaboration networks:  Average number of authors per paper ($\AvgAuthorsPerPaper$), average number of papers per author ($\AvgPapersPerAuthor$), and average number of co-authors per author ($\AvgCollabsPerAuthor$).}\label{tab:all-collabs-summary}
\centering
\begin{tabular}{lrrrrl}
\hline\hline
Scientific & No. of &  \multicolumn{3}{c}{\underline{Fundamental Statistics}} & Literature Reference\\
Discipline & Years  & $\AvgAuthorsPerPaper$ & $\AvgPapersPerAuthor$ & $\AvgCollabsPerAuthor$\\
\hline
\underline{International Research}\\
\quad Biomedical Research  & 40 & 3.75 &  6.40 &  18.10 & \citet{newman01a}\\
\quad High-energy Physics  & 27 & 8.96 & 11.60 & 173.00 & \citet{newman01a}\\
\quad CS                   & 10 & 2.22 & 2.55  &   3.59 & \citet{newman01a}\\
\underline{Philippine-based Research}\\
\quad Physics              &  5 & 3.16 & -    &  10.80 & \citet{villanueva09}\\
\quad Agriculture          &  4 & 3.81 & 1.39 &   2.70 & \citet{pabico09a}\\
\quad Various Fields       &  4 & 3.70 & 1.52 &   2.82 & \citet{pabico09b}\\
\quad ABE                  &  3 & 3.02 & 1.59 &   2.35 & \citet{pabico10}\\
\quad {\bf CS}             &  {\bf 9} & {\bf 2.68} & {\bf 1.42} & {\bf 3.58} \\
\hline\hline
\end{tabular}
\end{table*}

\section{Materials and Methods}

\subsection{Archive of Scientific Papers}

We have utilized the author information from 326~peer-reviewed papers presented during the 2000 to 2010 PCSC~\citep{pcsc03,pcsc04,pcsc05,pcsc06,pcsc07,pcsc08,pcsc09,pcsc10}. The papers presented each year are archived electronically in CDROM format, which is distributed to PCSC participants and paper presentors during the conference. The CDROM contains papers that are usually in portable document format (PDF) and comes with a table of contents that is also in PDF. An easily parseable hypertext markup language (HTML) format of the archive is also accessible from the website of the Computing Society of the Philippines~\citep{csp}.  

Table~\ref{tab:pcsc-archives} summarizes the particulars of various PCSC such as their respective proceedings, the number of papers presented, and the number of authors who wrote the papers during each year. The number of papers and authors during the 2000 PCSC were closed to the annual average, respectively. Both counts increase steadily in the earlier 4--year span from 2003 to 2006. When the PCSC was held in Boracay in 2007, both the number of authors and papers dropped considerably. However, both counts gain momentum and increase considerably in the recent 4--year span from 2007 to 2010. The 2010 PCSC has received a record number of paper submissions, and thus reflects the record-breaking number of papers accepted and presented, as well as the number of authors who wrote the papers. The 9--year PCSC has attracted an annual average of 36~papers and 83~authors.

\begin{table*}[htb]
\caption{Basic information about the 2000 to 2010 Philippine Computing Science Congress: Year and site each held, number of papers presented, number of authors, and proceedings reference.}\label{tab:pcsc-archives}
\centering\begin{tabular}{clccll}
\hline\hline
Year & PCSC Site & Number of & Number of &  Proceedings & Remarks\\
     &           & Papers    & Authors   &  Reference   & \\
\hline
2000 & De La Salle University&  35  & 78 & \citet{pcsc00} & POSTERS\\
2001 & MSU-IIT               &  -   & -  & - & Data not available\\
2002 & -                     &  -   & -  & - & Not held\\
2003 & Philippine Science HS &  15  & 31 & \citet{pcsc03}\\
2004 & UP Los Ba\~nos        &  29  & 61 & \citet{pcsc04}\\
2005 & University of Cebu    &  33  & 80 & \citet{pcsc05}\\
2006 & Ateneo de Manila      &  38  & 101& \citet{pcsc06}\\
2007 & Boracay Island        &  33  & 74 & \citet{pcsc07}\\
2008 & UP Diliman            &  37  & 76 & \citet{pcsc08}\\
2009 & Silliman University   &  41  & 93 & \citet{pcsc09} & RIPS\\
2010 & Ateneo de Davao       &  61  & 148& \citet{pcsc10} & RIPS and POSTERS\\\hline
\multicolumn{2}{r}{\bf Average} &  36  & 83 \\
\hline\hline
\end{tabular}
\end{table*}

In this study, we considered a scientific paper as either a keynote paper, a plenary (invited) paper, a tutorial paper, or a contributed paper. These paper types are present in all PCSC with the exception of the first year and the latest two years. In 2000 PCSC, a poster paper session (POSTERS) was included and the 2000 archive includes these paper type. In PCSC 2009, the research-in-progress session (RIPS) was instituted. RIPS allows the oral presentation of papers that are usually authored by undergraduate students and are categorized by the paper review panelists as incomplete or {\em in progress} but are already worthy of oral presentation. The PCSC 2009 archive, however, did not label whether the paper was RIPS or not. Thus, we assumed here that the 2009 PCSC archive does not include the RIPS. In PCSC 2010, POSTERS was reinstituted. Both RIPS and POSTERS papers are included in the 2010 PCSC archive. However, we did not include these papers in our study because as of this writing, the author information is incomplete for papers with more than one author. 

In our analysis of the co-authorship patterns, we considered an archive $\Papers = \{P_1, P_2, \dots, P_N\}$ of $N$~scientific papers, with each paper~$P_i$ having a list $\Authors_i = \{A_j|A_j\in \Authors, |\Authors_i|=M_i\}$ of $M_i$~authors. From the author information in~$\Papers$, we created a database of distinct authors $\Authors=\union_{i=1}^N\Authors_i$, such that $M=|\Authors|\leq\sum_{i=1}^NM_i$. We note here that $M=\sum_{i=1}^NM_i$ implies $\intersection_{i=1}^N\Authors_i=\emptyset$, which means that all authors have written exactly one paper. Our results show that this is not the case in Philippine CS research.

\subsection{Building the Paper-Author Bipartite Graph}

Given~$\Papers$ and~$\Authors$, we built the paper--author bipartite graph $\PAG = (\Papers\union\Authors, \E)$, where $\E=\{(i,j)|P_i\in \Papers, A_j\in \Authors\}$. For each paper~$P_i$, we created a bipartite subgraph (sub-bigraph) $\PAGS_i$ composed of a type--$P$ vertex labeled~$P_i$, and $M_i$~type--$A$ vertices with the respective labels as in $\Authors_i$. We then created edges in $\PAGS_i$ by connecting the type--$P$ vertex with all the $M_i$~type--$A$ vertices. The $i$th sub-bigraph induced by $P_i$ represents the one-to-many relationship between the $i$th paper and its $M_i$~authors. We then connected all $N$~sub-bigraphs via each sub-bigraph's common type--$A$ vertices. The resulting graph $\bigcup_{i=1}^N \PAGS_i$ is the paper--author bipartite graph $\PAG$. Intuitively, $\PAG$ may be built with a time complexity of $O(N\times M)$ but we reduced this to $O(N\times \log M)$ by using a balanced binary tree structure for~$\Authors$.

\begin{figure*}[htb]
\centering\epsfig{file=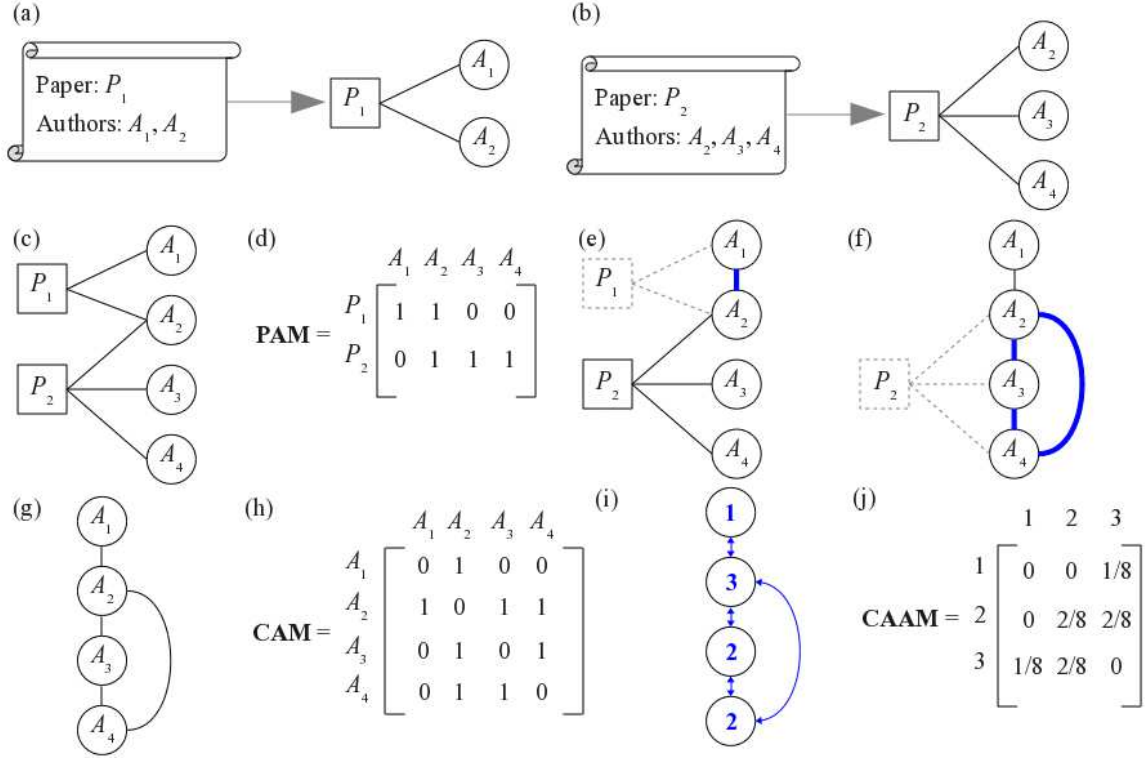, width=6in}
\centering\caption{The process flow for building the paper--author bipartite graph $\PAG$ and the paper--author matrix $\PAM$ using an archive with two hypothetical papers~$P_1$ and~$P_2$: (a)~The hypothetical paper~$P_1$ and its corresponding sub-bigraph; (b)~The hypothetical paper~$P_2$ and the sub-bigraph induced by it; (c)~The resulting $\PAG=\PAGS_1\union\PAGS_2$; and (d)~The equivalent $\PAM$. The process flow for building the co-authorship graph $\CAG$ and the co-authorship matrix $\CAM$: (e)~Deleting vertex~$P_1$ and edges~$(1,1)$ and~$(1,2)$, and creating the completely connected subgraph~$\CAGS_1$; (f)~Deleting vertex~$P_2$ and edges $(2,2)$, $(2,3)$, and~$(2,4)$, and creating the fully-connected subgraph~$\CAGS_2$; (g)~The resulting~$\CAG=\CAGS_1\union\CAGS_2$; and (h)~The equivalent $\CAM$. The process flow for transforming $\CAG$ into $\CAAG$ and the corresponding mixing matrix $\CAAM$: (i)~The mixing network when $\tau=\Delta$; and (j)~The resulting $\CAAM$. In the visualization of the different graphs, square vertices represent papers while circle vertices represent authors.}\label{fig:flowchart}
\end{figure*}

Figure~\ref{fig:flowchart}(a--c) shows how the $\PAG$ was created for a hypothetical paper archive~$\Papers$ composed of two papers~$P_1$ and~$P_2$ written by authors $A_1$, $A_2$, $A_3$, and $A_4$. In this scenario, $P_1$~was co-authored by~$A_1$ and~$A_2$, while~$P_2$ was jointly written by~$A_2$, $A_3$, and~$A_4$. In both papers, $A_2$~was the common author. Separately, the sub-bigraph induced by $P_1$ is $\PAGS_1 = (\{P_1, A_1, A_2\}, \{(1,1), (1,2)\})$, while the sub-bigraph induced by $P_2$ is $\PAGS_2=(\{P_2, A_2, A_3, A_4\}, \{(2,2), (2,3), (2,4)\})$. The sub-bigraphs $\PAGS_1$ and $\PAGS_2$ are connected through the common vertex~$A_2$ to create $\PAG = (\Papers\union \Authors, \E)$, where $\Papers=\{P_1, P_2\}$, $\Authors=\{A_1, A_2, A_3, A_4\}$, and $\E = \{(1,1), (1,2), (2,2), (2,3), (2,4)\}$.

\subsection{Building the Co-authorship Graph}

We built the co-authorship graph $\CAG$ from $\PAG$ as follows. For each vertex~$P_i\in \PAG$, we deleted all incedent edges to (or from)~$P_i$, as well as~$P_i$ itself, and created in its instead a complete subgraph $\CAGS_i = (\Authors_i, \E_i)$, where $\E_i = \{(j,k)| A_j,A_k\in \Authors_i, j\ne k\}$ and $|\E_i| = M_i(M_i-1)/2$ connecting all pairwise combinations of $A_j, A_k\in \Authors_i$, $j\ne k$. The fully-connected subgraph~$\CAGS_i$ represents the co-authorship graph of authors who co-wrote the $i$th paper~$P_i$. The resulting graph $\CAG = \union_{i=1}^N \CAGS_i$ is the co-authorship graph of CS researchers in the Philippines. Because some authors have not collaborated, some vertices $A_i\in \CAG$ are not connected to any of the other vertices $A_j\in \CAG$.  

Figure~\ref{fig:flowchart}(e--g) shows the flow diagram of the procedure on how~$\CAG$ was created from the hypothetical example mentioned above. The co-authorship subgraph induced by~$P_1$ is $\CAGS_1=(\{A_1, A_2\}, \{(1,2)\})$, while the co-authorship subgraph induced by~$P_2$ is $\CAGS_2=(\{A_2, A_3, A_4\}, \{(2,3), (2,4), (3,4)\})$. The subgraphs~$\CAGS_1$ and~$\CAGS_2$ are connected through the common vertex~$A_2$ to create the co-authorship graph $\CAG(\Authors,\E_c)$, where $\Authors=\{A_1, A_2, A_3, A_4\}$ and $\E_c = \{(1,2), (2,3), (2,4), (3,4)\}$. 

In building~$\CAG$, we adopted the same assumptions made by~\citet{newman01a}: (1) That all pairs of authors~$A_i$ and~$A_j$, $\forall i\neq j$, who have written a paper together are genuinely acquainted with one another; and (2) That the co-authorship graph~$\CAG$ reflects a genuine professional interaction between Filipino computer scientists. 

\subsection{Computing for node degrees}

From $\PAG$, we can infer an $N\times M$ matrix $\PAM$ that mathematically represents the adjacency of $\PAG$. Each matrix element $\PAM_{i,j}=1$ if the $i$th paper is written or co-written by the $j$th author. Otherwise, $\PAM_{i,j}=0$. The $\PAM$ of the hypothetical~$\PAG$ discussed above is shown in Figure~\ref{fig:flowchart}(d). It is interesting to note that $\PAM_{i,j}\not>1$ as no distinct author name appears more than once in the author line of a paper.

The degree $\degreeP_i$ of the $i$th $P$--type vertex~$P_i$ represents the number of authors that wrote paper~$P_i$. Conversely, the degree $\degreeA_j$ of the $j$th $A$--type vertex $A_j$ represents the number of papers that author~$A_j$ wrote. The vertex degrees $\degreeP_i$ and $\degreeA_i$ can be computed using $\PAM$ as shown in Equations~\ref{eqn:paper-degree} and~\ref{eqn:author-degree}, respectively. We can use the vertex degrees to compute for the minimum~($\MinAuthorsPerPaper$), average~($\AvgAuthorsPerPaper$) and maximum~($\MaxAuthorsPerPaper$) number of authors per paper (Equations~\ref{eqn:alpha-min} to~\ref{eqn:alpha-max}), as well as the minimum~($\MinPapersPerAuthor$), average~($\AvgPapersPerAuthor$) and maximum~($\MaxPapersPerAuthor$) number of papers per author (Equations~\ref{eqn:phi-min} to~\ref{eqn:phi-max}). 
The degree $\degreeP_i$ of the $i$th $P$--type vertex~$P_i$ represents the number of authors that wrote paper~$P_i$. Conversely, the degree $\degreeA_j$ of the $j$th $A$--type vertex $A_j$ represents the number of papers that author~$A_j$ wrote. The vertex degrees $\degreeP_i$ and $\degreeA_i$ can be computed using the matrix $\PAM$ as shown in Equations~\ref{eqn:paper-degree} and~\ref{eqn:author-degree}, respectively. We can use the vertex degrees to compute for the minimum~($\MinAuthorsPerPaper$), average~($\AvgAuthorsPerPaper$) and maximum~($\MaxAuthorsPerPaper$) number of authors per paper (Equations~\ref{eqn:alpha-min} to~\ref{eqn:alpha-max}), as well as the minimum~($\MinPapersPerAuthor$), average~($\AvgPapersPerAuthor$) and maximum~($\MaxPapersPerAuthor$) number of papers per author (Equations~\ref{eqn:phi-min} to~\ref{eqn:phi-max}). 

In the hypothetical archive discussed above, $\degreeP_1=2$~while~$\degreeP_2=3$. Conversely, $\degreeA_1 = 1$, $\degreeA_2 = 2$, $\degreeA_3=1$, and $\degreeA_4 = 1$. $\MinAuthorsPerPaper=2$, $\AvgAuthorsPerPaper=2.5$, and $\MaxAuthorsPerPaper=3$. Similarly, $\MinPapersPerAuthor=1$, $\AvgPapersPerAuthor=1.25$, and $\MaxPapersPerAuthor=2$.

From~$\CAG$, we can infer an $M\times M$ diagonally symmetric co-authorship matrix~$\CAM$ that mathematically represents ties between the $M$~scientists. Each matrix element $\CAM_{j,k}=\CAM_{k,j}=1$  if and only if author $A_j$ has collaborated with author $A_k$ with at least one paper. Since collaboration is a symmetric relation, $\CAM_{j,k}=1$ implies $\CAM_{k,j}=1$, which means that author $A_k$ collaborates with author $A_j$ in response. Without losing generality, we set all diagonal elements $\CAM_{j,j}=0$. If $A_j$ has not collaborated with $A_k$, then $\CAM_{j,k}=\CAM_{k,j}=0$. Figure~\ref{fig:flowchart}(h) shows the computed~$\CAM$ of the hypothetical~$\CAG$. Using~$\CAM$, the vertex degree~$\degreeC_i$ of the $i$th author, which reflects the number of co-authors~$A_i$ has, is computed as shown in Equation~\ref{eqn:vertex-degree}, while the minimum~$\MinCollabsPerAuthor$, average~$\AvgCollabsPerAuthor$, and maximum~$\MaxCollabsPerAuthor$ number of co-authors are respectively computed as in Equations~\ref{eqn:zeta-min} to~\ref{eqn:zeta-max}. 
\begin{eqnarray}
  \degreeP_i &=& \sum_{j=1}^M \PAM_{i,j}\label{eqn:paper-degree}\\
  \degreeA_j &=& \sum_{i=1}^N \PAM_{i,j}\label{eqn:author-degree}\\
  \degreeC_i &=& \sum_{j=1}^M \CAM_{j,i}\label{eqn:vertex-degree}\\
  \MinAuthorsPerPaper  &=&\min_{i=1}^N \degreeP_i\label{eqn:alpha-min}\\
  \AvgAuthorsPerPaper  &=&\frac{\sum_{i=1}^N \degreeP_i}{N}\label{eqn:alpha-avg}\\
  \MaxAuthorsPerPaper  &=&\max_{i=1}^N \degreeP_i\label{eqn:alpha-max}\\
  \MinPapersPerAuthor  &=&\min_{j=1}^M \degreeA_j\label{eqn:phi-min}\\
  \AvgPapersPerAuthor  &=&\frac{\sum_{j=1}^M \degreeA_j}{M}\label{eqn:phi-avg}\\
  \MaxPapersPerAuthor  &=&\max_{j=1}^M \degreeA_j\label{eqn:phi-max}\\
  \MinCollabsPerAuthor &=&\min_{i=1}^M \degreeC_i\label{eqn:zeta-min}\\
  \AvgCollabsPerAuthor &=&\frac{\sum_{i=1}^M \degreeC_i}{M}\label{eqn:zeta-avg}\\
  \MaxCollabsPerAuthor &=&\max_{i=1}^M \degreeC_i\label{eqn:zeta-max}
\end{eqnarray}

\subsection{Degree Distributions in $\PAG$ and $\CAG$}

The frequency distribution~$\rho(\degree)$ of a vertex degree~$\degree$ is a graph--based quantity that has been much studied and applied recently for various co-authorship graphs~\citep{newman01a,newman01b,bird09} and social networks~\citep{pabico08, arevalo09}. It provides the frequency that a randomly selected vertex has $\degree$~edges (or degrees). Graphs with high-degree yet low cardinality vertices have long-tailed~$\rho(\degree)$ and are called {\em scale-free} graphs. Such graphs follow the power law distribution (Equation~\ref{eqn:power-law}) and oftentimes model the relationships of naturally occuring entities, such as that of proteins and their interactions~\citep{salwinski04}. We hypothesized that~$\PAG$ and~$\CAG$ are scale-free and thus their respective $\rho$ follow a power-law. To test this hypothesis, we fitted a power law line each on~$\rho(\degreeP)$, $\rho(\degreeA)$, and $\rho(\degreeC)$ and statistically tested the power to be significantly different from zero at $\alpha=0.05$ (where $\alpha$ is taken as the probability of the two-tailed alternative greater than the test statistics). The power law distribution is statistically estimated by the frequency~$y$ in Equation~\ref{eqn:linest} and involves the vertex degree~$\degree$, a constant~$c$, and the power~$\power$, which is also known as the fractal dimension~\citep{kim07}. We estimated the values of~$c$ and~$\power$ by using a linear regression analysis in the power law's linear form (Equation~\ref{eqn:linest}).

\begin{eqnarray}
  y &=& c \degree^\power\label{eqn:power-law}\\
  \log y &=& \log c + \power \log \degree\label{eqn:linest}
\end{eqnarray}

\subsection{Productivity and collaboration}

An author $A_i \in \Authors$ has an inherent vector of valued attributes $(\tau_1, \tau_2)$, wherein in this research we set $\tau_1=\PapersPerAuthor$ and $\tau_2=\CollabsPerAuthor$. we hypothesized that $\PapersPerAuthor$ and $\CollabsPerAuthor$ have a high positive correlation such that authors who are productive, as measured by their high $\PapersPerAuthor$, are also those who have high number of memberships in various collaboration efforts, as measured by their high $\CollabsPerAuthor$. High positive correlation would also mean that authors who are less productive (i.e., low $\PapersPerAuthor$) are those who write alone or their number of collaborators is relatively small (i.e., low $\CollabsPerAuthor$). We tested the hypothesis by estimating the Pearson correlation $r$ and statistically testing it against zero (i.e., we hypothesize that $r\not=0$). We utilized the Pearson statistics because the causality relation between $\PapersPerAuthor$ and $\CollabsPerAuthor$ was not established (i.e., we do not know whether $\PapersPerAuthor$ causes $\CollabsPerAuthor$, or {\em vice versa}, or whether such relation exists at all). 

\subsection{Assortativity in $\CAG$}

Given an attribute~$\tau$ of a vertex, the assortativity~$r$ of a graph is the tendency of vertices to be connected to like vertices~\citep{newman06}, such that there are more edges between vertices with high $\tau$ values than between a high--$\tau$ vertex and a low--$\tau$ vertex. We started its computation by relabeling each vertex $A_i\in \CAG(\Authors,\E_c)$ by its~$\tau$, and converting all undirected edges in $\E$ to bidirectional edges to create $\E_d$. The resulting graph~$\CAAG(\Authors',\E_d)$, where~$\Authors'$ is just the relabeled vertices in~$\Authors$, and $|\E_d|=2\times |\E_c|$. We used a {\em mixing matrix}~$\CAAM$, where each matrix element~$\CAAM_{i,j}$ represents the fraction of all edges in~$\CAAG$ that start at $a_i$ and end at $a_j$, such that $\sum_{i,j} \CAAM_{i,j} = 1$. Let $f_i$ be the fraction of all edges in~$\CAAG$ that are incident to $a_i$, thus $f_i = \sum_j \CAAM_{i,j}$. The assortativity~$r$ can be approximated by the Pearson correlation coefficient discussed by~\citet{newman06} and subsequently used by~\citet{bird09}. {\em Assortativity} is when all vertices in $\CAG$ are connected only to vertices with similar~$\tau$ (i.e., $r>0$). {\em Dissortativity} (or negative assortativity~$r<0$) is when high--$\tau$ vertices are only connected to low--$\tau$ ones. Using the degree~$\degree$ as $\tau$, Figure~\ref{fig:flowchart}(i--j) shows how the~$\CAG$ of the hypothetical example discussed above was transformed into~$\CAAG$, as well as how the~$\CAAM_{i,j}$ was computed. In this paper, we independently used $\PapersPerAuthor$ and $\CollabsPerAuthor$ as $\tau$ to separately discover the general preference of CS researchers in choosing a collaborator in terms of the collaborator's $\PapersPerAuthor$ and $\CollabsPerAuthor$, respectively.

\section{Results and Discussion}

\subsection{The PCSC Paper Archive}

For this study, we utilized the archive~$\Papers$ of papers presented during the 2000 to 2010 PCSC to infer the authorship patterns of Filipino computer scientists. The total number of papers presented in these conventions is $N=326$, while the number of authors is $M=605$. As pointed out by~\citet{newman01a}, one particular issue that we were concerned about was the number of names~$L$ that appear in~$\Papers$, which clearly identifies distinct authors. This is because~$L$ is not necessarily the same as~$M$. For example, author~$A_i$ may format his name differently on different papers, such that the names {\em Juan dela Cruz}, {\em Dela Cruz, Juan}, and {\em J. dela Cruz} could all belong to him. This scenario would mean that $M=3$, but in fact $L=1$. On the contrary, two distinct authors~$A_i$ and~$A_j$ may have the same name, such that the name {\em Maria Maquiling} could belong to both. This means that $M=1$, while in fact $L=2$. This apparent name ambiguity problem has already been given approximate solutions by various techniques~\citep{huang06,cucerzan07,wang08,torvik09} that use additional information found in the papers, such as the names of the authors' respective home institutions and their subdisciplines. However, we could not use these additional information because there are authors who belong to more than one institution, and due to multi-specialty research collaborations, they could publish in other subdisciplines. Further, the author information in~$\Papers$ rarely includes the subdisciplines. In order to solve these issues, we performed our analysis using the author's surname and first and second names' initials, knowing full well that we may be overestimating the true value of~$M$. In this regard, having $L\geq M$ in this research may give us a guarantee that our results provide the respective upper bounds of the patterns.

\begin{figure}[hbt]
\centering\epsfig{file=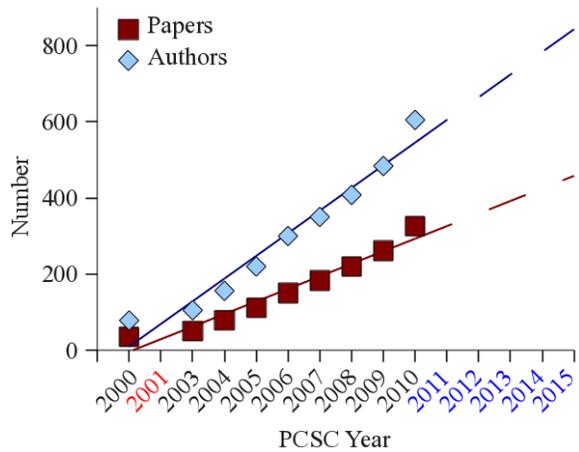,width=3in}
\caption{Annual trend of the cumulative number of papers (red square) and authors (blue diamond). The red solid and dashed lines, respectively, represent the regression and the 5-year extrapolation line of the yearly cumulative number of papers (slope$=32.99$, $R^2=0.95$). The blue solid and dashed lines, respectively, represent the regression and the 5-year extrapolation line of the yearly cumulative number of authors (slope$=59.54$, $R^2=0.96$). (This figure is in color in the digital format of this paper.)}\label{fig:annual-trend}
\end{figure}

Figure~\ref{fig:annual-trend} shows the annual trend of cumulative number of authors and papers presented in the 9--year PCSC. Based on simple regression analysis, we found out that PCSC has attracted about 60~new authors per year who helped co-write about 33~new papers annually. After extrapolating these lines to 5~years into the future, we can see that in 2015 the number of distinct authors that will be contributing to PCSC will reach to~843 while the number of papers that will be contributed will reach to~458.

\subsection{Inferences from~$\PAG$}\label{sec:paper-author}

\begin{figure*}[htb]
\centering\epsfig{file=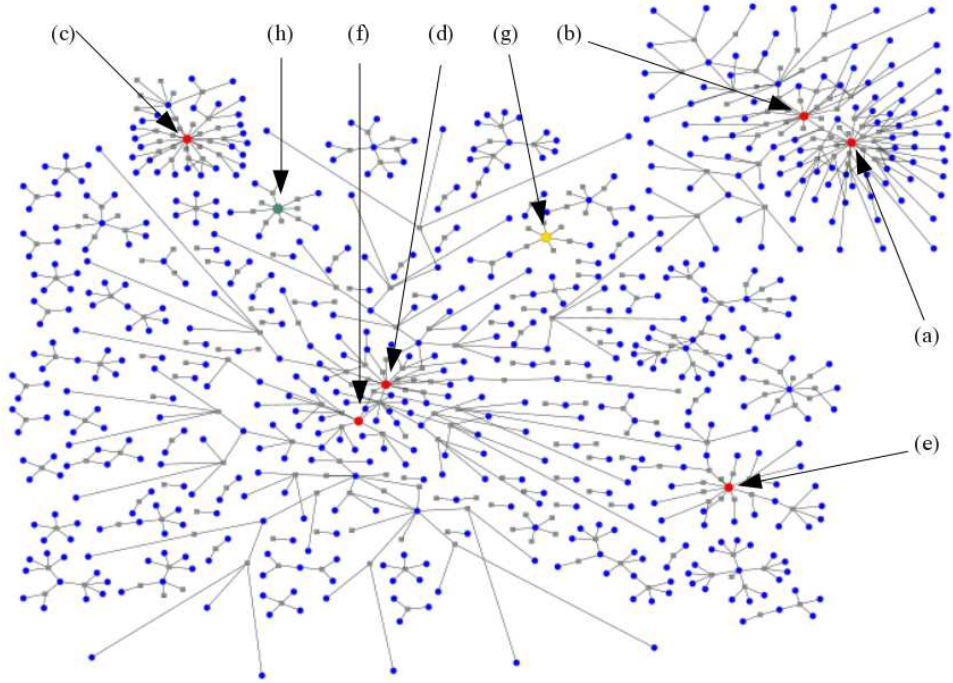,width=5in}
\caption{The paper--author bigraph~$\PAG$ drawn with the graph visualization algorithm by~\citet{kamada89}. In this visualization, colored circles represent authors while gray squares represent papers. The labels correspond to some identified authors with the most number of papers: (a)~PC~Naval, (b)~HN~Adorna, (c)~RP~Salda\~na, (d)~R~Sison, (e)~REO~Roxas, and (f)~D~Cheng. In addition, (g)~is this journal's editor-in-chief EA~Albacea, while (h)~is this paper's author. (This figure is in color in the digital format of this paper.)}\label{fig:actual-PAG}
\end{figure*}

Table~\ref{tab:simple-stat} summarizes the values inferred from~$\PAG$. On the average, the CS authors in the Philippines have writen about 1.42~papers, while papers were written by an average of 2.64~authors. The Filipino authors have collaborated, on the average, with 3.70 other authors. We have shown the comparison of these simple statistics with other various national and international research co-authorship graphs (Table~\ref{tab:all-collabs-summary}). As inferred also from $\PAG$, we have identified the top five researchers with the most number of papers in the archive: PC~Naval (20 papers), RP~Sala\~na~(19), HN~Adorna~(16), RC~Sison~(15), and REO~Roxas and JDL~Caro (10 each). We have annotated the vertices in Figures ~\ref{fig:actual-PAG} and~\ref{fig:collab-net2} to visualize the respective relative positions of these authors in $\PAG$ and $\CAG$. 

\begin{table*}[htb]
\caption{Values of inferred statistics from~$\PAG$ and $\CAG$: Minimum, average and maximum number of authors per paper ($\AuthorsPerPaper$), number of papers per author ($\PapersPerAuthor$), and number of collaborators per author ($\CollabsPerAuthor$); As well as the respective degree distribution's power law coefficients ($\power$) and the corresponding statistics ($R^2$).}\label{tab:simple-stat}
\centering\begin{tabular}{lccccc}
\hline\hline
Statistics & Mininum & Average & Maximum & \multicolumn{2}{c}{\underline{Degree Distribution}}\\
           &         &         &         & $\power$  & $R^2$\\\hline
Number of authors per paper & $\MinAuthorsPerPaper=1$ & $\AvgAuthorsPerPaper=2.64$ & $\MaxAuthorsPerPaper=13$ & $-2.04$ & $0.71$\\
Number of papers per author & $\MinPapersPerAuthor=1$ & $\AvgPapersPerAuthor=1.42$ & $\MaxPapersPerAuthor=20$ & $-1.88$ & $0.83$\\
Number of collaborators per author & $\MinCollabsPerAuthor=0$ & $\AvgCollabsPerAuthor=3.70$ & $\MaxCollabsPerAuthor=54$ & $-1.65$ & $0.80$\\
\hline\hline
\end{tabular}
\end{table*}

\subsection{Number of authors per paper}

The Filipino CS research papers have been written on the average by 2.64 authors, which is lower compared to that of the ABE ($\AvgAuthorsPerPaper=3.02$), the agricultural science ($\AvgAuthorsPerPaper=3.81$) and NAST sciences ($\AvgAuthorsPerPaper=3.70$) in the country. This means that in the Philippines, creating new scientific information requires less number of authors in CS than in other disciplines. In the international co-authorship graphs, more authors are needed to write new information in the field of biomedical research ($\AvgAuthorsPerPaper=3.75$), and significantly more authors in the high-energy physics ($\AvgAuthorsPerPaper=8.96$). However, the Filipino CS research papers needed more authors on the average compared to that in the international CS's ($\AvgAuthorsPerPaper=2.22$).

\subsection{Number of papers per author}

 On the average, the Filipino CS researchers have written less papers ($\AvgPapersPerAuthor=1.42$) than their ABE ($\AvgPapersPerAuthor=1.59$) and NAST ($\AvgPapersPerAuthor=1.52$) counterparts, but more than the agricultural ($\AvgPapersPerAuthor=1.39$) scientists in the country. However, the average scientific productivity of Filipino computer scientists, measured by the number papers written per author, still falls behind the international averages. The international biomedical researchers, high-energy physicists, and computer scientists have respectively written an average of 6.4, 11.6, and 2.55 papers. 

\subsection{Inferences from~$\CAG$}\label{sec:co-authorship}

\begin{figure*}[ht]
\centering\epsfig{file=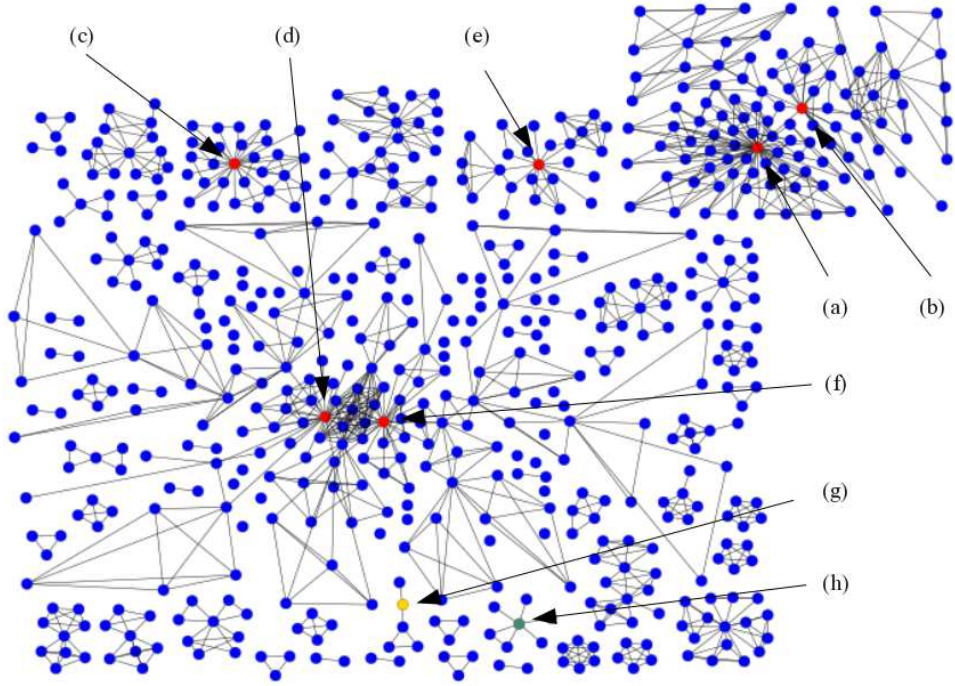,width=5in}
\caption{The co-authorshop graph~$\CAG$ of Filipino computer scientists is a {\em sociogram} that shows the professional relations between scientists involved in scientific research. The sociogram was drawn using the two-dimensional force-directed algorithm by~\citet{kamada89}. The labels correspond to some identified authors believed to be central persons in their respective subgraphs. (a)~P~Naval, (b)~HN~Adorna, (c)~RP~Salda\~na, (d)~R~Sison, (e)~REO~Roxas, and (f)~D~Cheng are the top researchers with the most number of collaborators. (g)~This journal's editor-in-chief EA~Albacea with his own collaboration subgraph. (h)~This paper's author, who is also a central person in his own, although small, subgraph.  (This figure is in color in the digital format of this paper.)}\label{fig:collab-net2}
\end{figure*}

Figure~\ref{fig:collab-net2} presents a visualization of the co-authorship graph~$\CAG$ created from the papers in~$\Papers$. In this visualization, it can easily be seen that the graph of CS research co-authorship in the Philippines is composed of disconnected subgraphs. We have found out that authors in each of the subgraphs belong to the same institution. This means that CS authors collaborate only to authors who belong to the same institution, and that cross--institution collaborations do not exist yet in the Philippines setting. It is understandable, however, that not much nationally important computational problems exist, or have been identified, nowadays to bring researchers from several institutions together to solve a common problem. We have also identified and labeled some {\em central} authors in some of the subgraphs. We have identified the top five scientists with the most number of collaborators namely, PC~Naval with 54~collaborators, RC~Sison with 30, D~Cheng with 29, RP~Salda\~na with 25, and HN~Adorna with~20. We believed that these authors, together with those whom we identified with the most number of papers, are the {\em central} scientists in their respective subgraphs. By central we mean the most influential person among the connected authors in the subgraph.

\subsection{Number of collaborators per author}

In the area of collaborative research, the Filipino CS researchers have collaborated with more other researchers ($\AvgCollabsPerAuthor=3.70$) compared to that of their ABE ($\AvgCollabsPerAuthor=2.35$), agriculture ($\AvgCollabsPerAuthor=2.70$) and NAST ($\AvgCollabsPerAuthor=2.82$) counterparts in the country. The Filipino physicists, however, have more collaborators ($\AvgCollabsPerAuthor=10.80$) than the computer scientists. Similarly, the international scientists have collaborated significantly more compared to the Filipino computer scientists, with the biomedical researchers and high-energy physicists having an average collaborators of 18.1 and 173, respectively. This seemingly high number of collaborators in high-energy physics is actually achievable, as pointed out by~\citet{newman01a}, because of the significantly higher average number of authors per paper in their community ($\AvgCollabsPerAuthor=8.96$). Thus, the {\em mega--collaboration} average of 173 is actually just a product of their high $\AuthorsPerPaper$. The Filipino CS researchers, on the other hand, have collaborated with almost the same number of collaborators as that of the international counterparts~($\AvgCollabsPerAuthor=3.59$).  

\subsection{Degree Distributions}
\begin{figure*}[htb]
\centering\epsfig{file=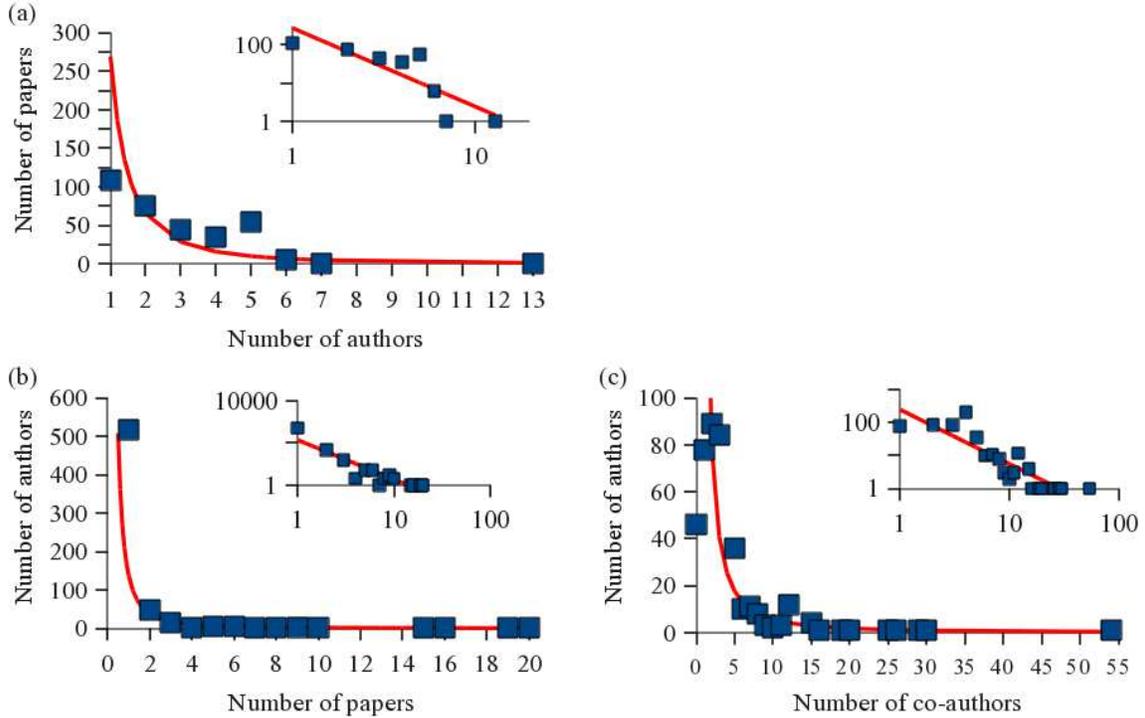,width=6in}
\caption{The vertex degree distributions follow the power law. Blue squares mean frequency of the vertex degree while the red dashed line is the power law fit. INSET: The same scatter and line plots in log-log scale. (a) $\rho(\degreeP): y = 269.15^* (\degreeP)^{-2.04^*}, R^2=0.71$; (b) $\rho(\degreeA): y=138^* (\degreeA)^{-1.88^*}, R^2=0.83$; and (c) $\rho(\degreeC): y = 251.03^* (\degreeC)^{-1.65^*}, R^2=0.80$. *The estimated coefficients are significantly different from zero at $\alpha=0.01$ statistics. (This figure is in color in the digital format of this paper.)}\label{fig:degree-distributions}
\end{figure*}

Figures~\ref{fig:degree-distributions} shows the respective degree ($\degreeP$, $\degreeA$, and $\degreeC$) frequencies of the vertices in $\PAG$ and $\CAG$, each plotted in scatter (for raw data) and line (predicted) plots. Figure~\ref{fig:degree-distributions}(a) shows the scatter and predicted line plots of the frequency distribution of the number of authors per paper in normal and log-log scales. Here we see that the predicted line plots follow a power law form. The power law line that we we found has the form $y = 269.15 (\degreeP)^{-2.04}$ with $R^2=0.71$. Both coefficients $c=269.15$ and $\power=-2.04$ are significantly different from zero at 1\% statistics, respectively, confirming our hypothesis that $\rho(\degreeP)$ obeys a power law distribution. We did not include the distribution for $\degreeA=0$ because no paper could have been written by zero authors (i.e., no paper has a missing author information). 

Figure~\ref{fig:degree-distributions}(b) shows the scatter and predicted line plots of the frequency distribution of the number of papers per author. Here we see that the line plot follows a power law of the form $y=138 (\degreeA)^{-1.88}$ with $R^2=0.83$. The coefficients $c=138$ and $\power=-1.88$ are significantly different from zero at 1\% statistics. Here, we did not include the distribution for $\degreeP=0$ because the CS researchers who have written zero papers are not included in the archive. The distribution generally characterizes a high number of authors who wrote a small number of papers, and a small number of authors who wrote a very large number of papers. Thus, in CS research the Philippines, the number of highly productive researchers is a relatively small fraction of all Filipino CS scientists. The power $\power=-1.88$ is in close agreement with Lotka's {\em law of scientific productivity} found in an experiment in 1926 to be $\approx -2$~\citep{lotka26}, while the coefficient $c=138$ uniquely characterizes the scientific productivity of CS researchers in the Philippines.

Figure~\ref{fig:degree-distributions}(c) shows the degree frequency of the vertices in $\CAG$ with a power law line fit of the form $y = 251.03 (\degreeC)^{-1.65}$ having a $R^2=0.80$. We accept that the power law is the best model for~$\rho(\degreeC)$ because we have found both $c=251.033$ and $\power=-1.65$ to be significantly different from zero at 1\% statistics. A power law fit suggests that:
\begin{enumerate}
\item Only a few number of authors have the most number of co-authors in~$\CAG$. These authors act as information hubs in the co-authorship graph, and therefore has the potential control of information flow through the network. We deemed such authors as influential or {\em central}. We have already identified these {\em central} persons in \S~\ref{sec:paper-author} and \S~\ref{sec:co-authorship}.
\item The co-authorship of CS research in the country is scale invariant. This means that the properties of~$\CAG$ that we observed in this study, as well as the patterns of co-authorship and publication, will not change as much when the number of authors~$M$ increases. This makes~$\CAG$ a {\em scale-free} graph.
\end{enumerate}

\subsection{Correlation Between $\PapersPerAuthor$ and $\CollabsPerAuthor$}

\begin{figure*}[htb]
\centering\epsfig{file=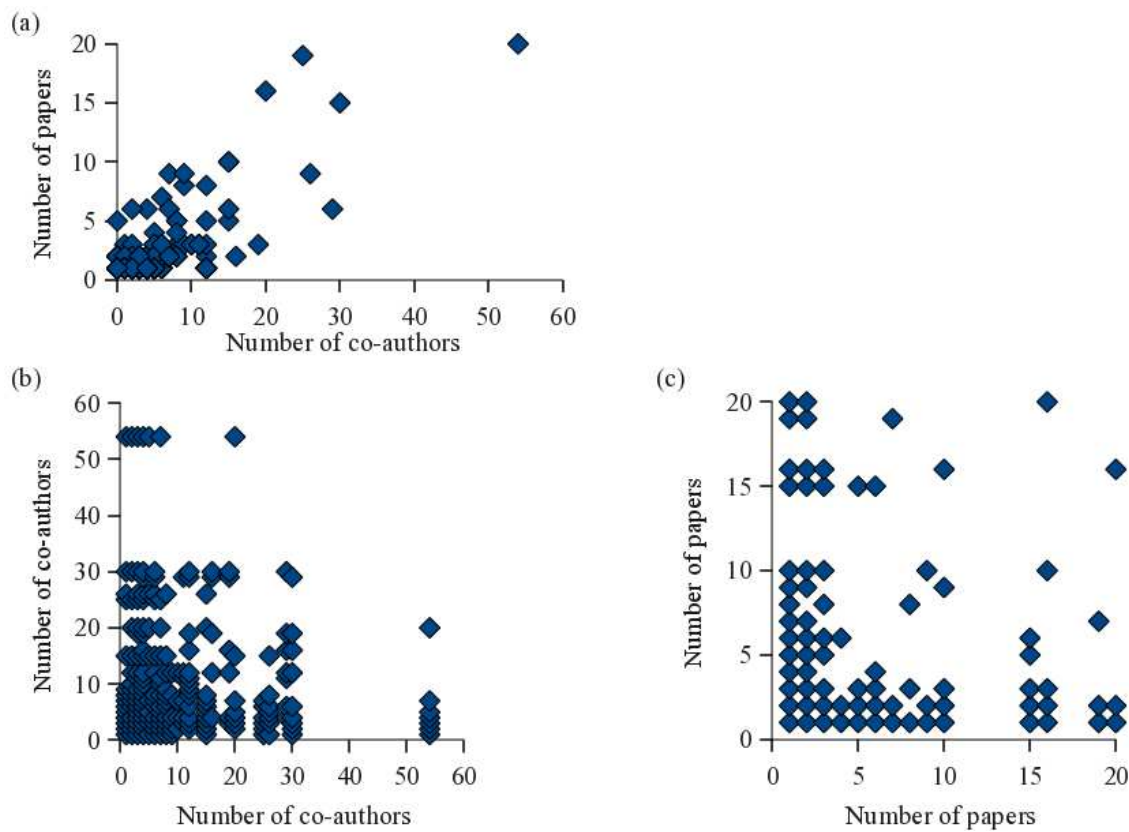,width=6in}
\caption{(a) Scatter plot between the number of papers and the number of collaborators of each author shows a positive correlation of $r=0.7425$; (b) Mixing plot between the number of papers per author shows a low negative correlation of $r=-0.1015$; and (c) Mixing plot between the number of co-authors per author shows a low negative correlation of $r=-0.0398$.}\label{fig:correlations}
\end{figure*}

Figure~\ref{fig:correlations}(a) shows the scatter plot between $\PapersPerAuthor$ and $\CollabsPerAuthor$. The scatter plot shows that they are positively correlated with $r=0.7425$. This suggests that the scientific productivity of the country's CS researchers, as measured by their number of papers, is correlated with the researchers' participation in collaborative research efforts, as measured by their number of co-authors. A highly productive scientist is most likely to have a high number of collaborators, and {\em vice versa}. This observation is particularly true in scientific publications because a large group of scientists has more manpower available for writing papers.

\subsection{Assortative mixing in $\CAG$}

Figure~\ref{fig:correlations}(b--c) shows the mixing plots for correlating the $\PapersPerAuthor$  and the $\CollabsPerAuthor$ of each researcher. These correlations quantify how a computer scientist chooses his collaborator based on the similarity or dissimilarity of his and the collaborator's attributes. Based on the Pearson correlation analysis, a computer scientist chooses a collaborator who has a dissimilar $\PapersPerAuthor$ ($-0.1015$) or $\CollabsPerAuthor$ ($r=-0.0398$) as he has. We expect that a computer scientist with a low $\PapersPerAuthor$ will most likely chooses a collaborator whose $\PapersPerAuthor$ is high. 

\section{Summary and Conclusion}

In this paper, we inferred two graphs $\PAG$ and $\CAG$ from the author information of CS papers in the country using various computational techniques. The graphs were based on publication data in various PCSC with 326 papers written by 605 authors. A large number of calculations were performed on the graphs, including the fundamental averages $\AvgAuthorsPerPaper=2.64$, $\AvgPapersPerAuthor=1.42$, and $\AvgCollabsPerAuthor=3.70$. The respective frequency distributions of these quantities follow a power law which suggests that most papers were written by scientists with a small number of collaborators, while few papers were authored by those with large number of collaborators. Specifically, the power $\power=-1.88$ of the frequency distribution for $\PapersPerAuthor$ closely agrees with Lotka's {\em law of scientific productivity}. The productivity of the scientists, as measured by $\PapersPerAuthor$, is positively correlated with the scientist's participation in a number of collaborative research efforts, as measured by $\CollabsPerAuthor$, suggesting that  highly productive scientists are more likely to have a high number of collaborators, and scientists with high number of collaborators are more likely to be highly productive. The assortativity tests show that scientists prefer to conduct collaborative research endeavors with scientists whose number of papers and collaborators are different from theirs. It is therefore reasonable to suppose that the scientific enterprise in the CS field in the Philippines will be significantly be given a boost if collaboration among scientists will be promoted (e.g., maybe through governmental policies and other programs).

The following efforts are already underway as extensions to this research endeavor:
\begin{enumerate}
\item Time study to measure the dynamics and evolution of $\CAG$. The current effort did not measure how the current $\CAG$ has evolved to what it is today. Thus, the extended study tests several hypotheses regarding the nature of the development of the $\CAG$, including the social phenomenon called {\em preferential attachment}. Preferential attachment, also known as the “rich gets richer” adage, is the tendency of new scientists to build collaborations with prolific scientists, and then later on seek more collaborations with other prolific ones. These tendencies make scientists with high number of papers to write more papers in a given time than others.
\item Development of a National Researcher Database System. Due to the inherent name  ambiguity encountered in the conduct of this research, it is recommended that a National Researcher Database System (NRDS) be developed. NRDS will keep track of the changes in names used by a researcher, and at the same time be a repository of scientific articles in the Philippines. The content of the repository may be used as the National Index of Scientific papers in the Philippines. With the NRDS, a citation network may also be inferred  to compute the impact factor, not only of the journals and proceedings, but also of the papers themselves. 
\end{enumerate}

\section{Acknowledgments}

This research effort is funded by the Institute of Computer Science through UPLBGF \#2326103 and UPLBFI \#2004987. HN~Adorna provided us a copy of the 2000 PCSC, as well as the history of PCSC, which are valuable information that we used in this research.

\bibliography{research-collaboration}
\bibliographystyle{plainnat}

\end{document}